\documentclass[aps,prd,preprint,showpacs,superscriptaddress,nofootinbib]{revtex4}

\usepackage{graphicx}

\begin{document}

\title{Correlative signatures of heavy Majorana neutrinos and
doubly-charged Higgs bosons at the Large Hadron Collider}

\author{Wei Chao}  \email{chaowei@ihep.ac.cn}
\affiliation{Institute of High Energy Physics, Chinese Academy of
Sciences, Beijing 100049, China}

\author{Zong-Guo Si}  \email{zgsi@sdu.edu.cn}
\affiliation{Department of Physics, Shandong University, Jinan,
Shandong 250100, China}

\author{Zhi-zhong Xing}  \email{xingzz@ihep.ac.cn}
\affiliation{Institute of High Energy Physics, Chinese Academy of
Sciences, Beijing 100049, China}

\author{Shun Zhou}  \email{zhoush@ihep.ac.cn}
\affiliation{Institute of High Energy Physics, Chinese Academy of
Sciences, Beijing 100049, China}

\date{\today}

\begin{abstract}
We explore an intriguing possibility to test the type-II seesaw
mechanism of neutrino mass generation at the Large Hadron Collider
(LHC). We show that the lepton-number-violating signatures of
heavy Majorana neutrinos $N^{}_i$ (for $i=1, 2, 3$) and
doubly-charged Higgs bosons $H^{\pm \pm}$ can be closely
correlated with each other in a class of TeV-scale type-II seesaw
models. Taking the minimal version of such models for example, we
calculate the cross sections of $pp \rightarrow l^\pm_\alpha
l^\pm_\beta X$ processes mediated separately by $N^{}_1$ and
$H^{\pm\pm}$, and illustrate their nontrivial correlation at the
LHC.
\end{abstract}


\maketitle

\section{Introduction}

The running of the Large Hadron Collider (LHC) in the coming years
will shed light on several fundamental problems in the Standard
Model (SM) and explore possible new physics beyond the SM
\cite{Quigg}. If new physics exists at the TeV scale and is
responsible for the electroweak symmetry breaking, it may also be
responsible for the origin of neutrino masses. The latter is a kind
of new physics which has been firmly established by a number of
neutrino oscillation experiments in the past decade \cite{PDG}.
Therefore, it is extremely interesting to see whether some deep
understanding of the neutrino mass generation and lepton number
(flavor) violation can be achieved at the energy frontier set by the
LHC.

A natural and attractive possibility of generating tiny neutrino
masses is to extend the SM by introducing three heavy right-handed
Majorana neutrinos \cite{Seesaw1} and (or) one Higgs triplet
\cite{Seesaw2}. The gauge-invariant Lagrangian relevant to lepton
masses can then be written as
\begin{eqnarray}
-{\cal L}^{}_{\rm lepton} &=& \overline{l^{}_{\rm L}} Y^{}_l H
E^{}_{\rm R} + \overline{l^{}_{\rm L}} Y^{}_\nu \tilde{H} N^{}_{\rm
R} + \frac{1}{2} \overline{N^{c}_{\rm R}} M^{}_{\rm R} N^{}_{\rm R}
+ \frac{1}{2} \overline{l^{}_{\rm L}} Y^{}_\Delta \Delta
i\sigma^{}_2 l^c_{\rm L} + {\rm h.c.} \; ,
\end{eqnarray}
where $\Delta$ is the Higgs triplet, and $M^{}_{\rm R}$ is the
mass matrix of right-handed Majorana neutrinos. After the
spontaneous gauge symmetry breaking, one obtains the mass matrices
$M^{}_l = Y^{}_lv/\sqrt{2}$, $M^{}_{\rm D} = Y^{}_\nu v/\sqrt{2}$
and $M^{}_{\rm L} = Y^{}_\Delta v^{}_\Delta$, where $\langle H
\rangle \equiv v/\sqrt{2}$ and $\langle \Delta \rangle \equiv
v^{}_\Delta$ are the vacuum expectation values (vev's) of the
neutral components of $H$ and $\Delta$, respectively. In the
leading-order approximation, the effective mass matrix for three
light neutrinos is given by $M^{}_\nu \approx M^{}_{\rm L} -
M^{}_{\rm D} M^{-1}_{\rm R} M^T_{\rm D}$. This is the so-called
type-II seesaw mechanism. If the Higgs triplet $\Delta$ is absent,
the small mass scale of $M^{}_\nu$ can be just attributed to the
large mass scale of $M^{}_{\rm R}$ (type-I seesaw \cite{Seesaw1}).
In the absence of heavy right-handed Majorana neutrinos, the small
mass scale of $M^{}_\nu$ implies that the mass scale of $M^{}_{\rm
L}$ must be equally small (triplet seesaw \cite{Seesaw2}). Another
interesting case is that both terms of $M^{}_\nu$ are important
and their significant cancellation gives rise to small neutrino
masses \cite{typeII}. Direct tests of such neutrino mass models
can be done at the LHC, provided they work at the TeV scale and
predict appreciable collider signatures.

The search for the Higgs triplet and heavy Majorana neutrinos at the
Tevatron and LHC has recently attracted a lot of attention. For
example, a model-independent analysis has been done in Ref.
\cite{Han} to probe the same-sign dilepton events induced by heavy
Majorana neutrinos via the most promising channel $q \bar{q}^\prime
\to \mu^\pm N^{}_i \to \mu^\pm \mu^\pm W^\mp$. These events signify
the lepton number violation and serve for a clean collider signature
of new physics beyond the SM \cite{Senjanovic}. In {\it realistic}
type-I seesaw models, however, the observability of $N^{}_i$
requires ${\cal O}(M^{}_{\rm R}) \lesssim 1$ TeV and ${\cal
O}(M^{}_{\rm D}/M^{}_{\rm R}) \gtrsim 10^{-3}$ together with an
unnatural cancellation condition $M^{}_{\rm D} M^{-1}_{\rm R}
M^T_{\rm D} \approx {\bf 0}$ \cite{Pilaftsis,Smirnov}. In a triplet
seesaw model, the production of the doubly-charged Higgs bosons
$H^{\pm\pm}$ only depends on their masses and has nothing to do with
the Yukawa coupling $Y^{}_\Delta$. It is therefore possible to
discover $H^{\pm\pm}$ of ${\cal O}(\lesssim 1)$ TeV at the LHC by
detecting $H^{\pm\pm} \rightarrow l^\pm_\alpha l^\pm_\beta$ decays
\cite{Raidal1,Han2,Chun,Han3}.

Different from those previous works, this paper aims to investigate
possible correlation between the collider signatures of heavy
Majorana neutrinos $N^{}_i$ and doubly-charged Higgs bosons
$H^{\pm\pm}$ in a class of realistic type-II seesaw models, in which
the smallness of $M^{}_\nu$ is ascribed to a significant but
incomplete cancellation between large $M^{}_{\rm L}$ and $M^{}_{\rm
D} M^{-1}_{\rm R} M^T_{\rm D}$ terms. We shall see that the
phenomenology of such a type-II seesaw model is much richer than
that of a type-I seesaw model or a triplet seesaw model at the TeV
scale. In particular, the non-unitarity of the light neutrino mixing
matrix can be closely related to the lepton-number-violating signals
of both $N^{}_i$ and $H^{\pm\pm}$ at the LHC. Taking the minimal
version of the type-II seesaw models (with only one heavy Majorana
neutrino $N^{}_1$ in addition to the Higgs triplet $\Delta$) for
example, we calculate the cross sections of $pp \rightarrow
l^\pm_\alpha l^\pm_\beta X$ mediated separately by $N^{}_1$ and
$H^{\pm\pm}$, and illustrate their nontrivial correlation at the
LHC. We arrive at some interesting and encouraging conclusions.

\section{The model}

If an $SU(2)^{}_{\rm L}$ Higgs triplet is introduced into the SM
\cite{Seesaw2}, the gauge-invariant potential can be written as
$V(H, \Delta)= V^{}_{\rm SM}(H) + \delta V$, where $V^{}_{\rm SM}(H)
= -\mu^2 H^\dagger H + \lambda \left(H^\dagger H\right)^2$ with $H
\equiv \left(\varphi^+, \varphi^0\right)^T$ being the SM Higgs
doublet, and
\begin{eqnarray}
\delta V = \frac{1}{2} M^2_\Delta {\rm Tr}\left(\Delta^\dagger
\Delta\right) - \left[\lambda^{}_\Delta M^{}_\Delta H^T i\sigma^{}_2
\Delta H + {\rm h.c.}\right] \;
\end{eqnarray}
with $\Delta$ being defined as
\begin{equation}
\Delta \equiv \left(\matrix{\xi^- & - \sqrt{2} ~ \xi^0 \cr \sqrt{2}
~ \xi^{--} & -\xi^-}\right) \; .
\end{equation}
When the neutral components of $H$ and $\Delta$ acquire their
vev's $v$ and $v^{}_\Delta$, respectively, the electroweak gauge
symmetry is spontaneously broken. The minimum of $V(H,\Delta)$ can
be achieved at $v = \mu/(\lambda - 2\lambda^2_\Delta)^{1/2}$ and
$v^{}_\Delta = \lambda^{}_\Delta v^2/M^{}_\Delta$, where the
dimensionless parameter $\lambda^{}_\Delta$ has been assumed to be
real. Note that $v^{}_\Delta$ may affect the masses of $W^\pm$ and
$Z^0$ in such a way that $\rho \equiv M^2_W/(M^2_Z \cos^2
\theta^{}_{\rm W}) = (v^2 + 2v^2_\Delta)/(v^2 + 4v^2_\Delta)$
holds. By using current data on the $\rho$-parameter \cite{PDG},
we get $\kappa \equiv \sqrt{2} ~v^{}_\Delta /v < 0.01$ and
$v^{}_\Delta < 2.5~{\rm GeV}$. There are totally seven physical
Higgs bosons in this model: doubly-charged $H^{++}$ and $H^{--}$,
singly-charged $H^+$ and $H^-$, neutral $A^0$ (CP-odd), and
neutral $h^0$ and $H^0$ (CP-even), where $h^0$ is the SM-like
Higgs boson. Except for $M^2_{h^0} \approx 2\mu^2$, we get a
quasi-degenerate mass spectrum for other scalars: $M^2_{H^{\pm
\pm}} = M^2_\Delta \approx M^2_{H^0}$, $M^2_{H^\pm} = M^2_\Delta
(1 + \kappa^2)$, and $M^2_{A^0} = M^2_\Delta (1 + 2\kappa^2)$. As
a consequence, the decay channels $H^{\pm \pm} \to W^\pm H^\pm$
and $H^{\pm \pm} \to H^\pm H^\pm$ are kinematically forbidden. The
production of $H^{\pm\pm}$ at the LHC is mainly through $q\bar{q}
\to \gamma^*, Z^* \to H^{++}H^{--}$ and $q\bar{q}^\prime \to W^*
\to H^{\pm\pm}H^\mp$ processes.

On the neutrino side, we are left with $M^{}_{\rm L}$, $M^{}_{\rm
D}$ and $M^{}_{\rm R}$ from Eq. (1) after the electroweak symmetry
breaking. They form a symmetric $6 \times 6$ matrix ${\cal M}$,
which can be diagonalized by a unitary transformation ${\cal
U}^\dagger {\cal M} {\cal U}^* = \widehat{\cal M}$. Explicitly,
\begin{eqnarray}
\left(\matrix{V & R \cr S & U}\right)^\dagger \left( \matrix{
M^{}_{\rm L} & M^{}_{\rm D} \cr M^T_{\rm D} & M^{}_{\rm R}}\right)
\left(\matrix{V & R \cr S & U}\right)^*  = \left( \matrix{
\widehat{m} & {\bf 0} \cr {\bf 0} & \widehat{M}}\right) ,
\end{eqnarray}
$\widehat{m} = {\rm Diag}\{m^{}_1, m^{}_2, m^{}_3\}$ and
$\widehat{M} = {\rm Diag}\{M^{}_1, M^{}_2, M^{}_3\}$ with $m^{}_i$
and $M^{}_i$ (for $i=1, 2, 3$) being the light and heavy Majorana
neutrino masses, respectively. In the mass basis, the leptonic
charged-current interactions can be written as
\begin{eqnarray}
-{\cal L}^{}_{\rm cc} = \frac{g}{\sqrt{2}} \left[
\overline{e^{}_{\rm L}} V \gamma^\mu \nu^{}_{\rm L} W^-_{\mu} +
\overline{e^{}_{\rm L}} R \gamma^\mu N^{}_{\rm L} W^-_\mu \right] +
{\rm h.c.} \; .
\end{eqnarray}
Note that $VV^\dagger + RR^\dagger = {\bf 1}$ holds due to the
unitarity of ${\cal U}$, and thus the neutrino mixing matrix $V$
itself must be non-unitary \cite{Xing}. The unitarity violation of
$V$ is characterized by $R$, which is responsible for the production
and decays of heavy Majorana neutrinos $N^{}_i$. In the
leading-order approximation, the effective mass matrix of three
light neutrinos is given by $M^{}_\nu \approx M^{}_{\rm L} -
M^{}_{\rm D} M^{-1}_{\rm R} M^T_{\rm D}$. Either $M^{}_{\rm L}$ or
$M^{}_{\rm D} M^{-1}_{\rm R} M^T_{\rm D}$ may dominate $M^{}_\nu$,
but here we focus on the third possibility: the smallness of
$M^{}_\nu$ arises from a significant cancellation between $M^{}_{\rm
L}$ and $M^{}_{\rm D} M^{-1}_{\rm R} M^T_{\rm D}$ in the case of
${\cal O}(M^{}_\nu) \ll {\cal O} (M^{}_{\rm L}) \sim {\cal
O}(M^{}_{\rm D} M^{-1}_{\rm R} M^T_{\rm D})$. We admit that a
substantial fine-tuning at the level of ${\cal O}(10^{-10})$, which
is comparable with the degree of fine-tuning for the {\it structural
cancellation} of $M^{}_{\rm D}$ and $M^{}_{\rm R}$ in the TeV-scale
type-I seesaw models \cite{Pilaftsis,Smirnov}, has to be required in
order to realize this kind of {\it global cancellation}
\footnote{In our model, ${\cal O} (M^{}_{\rm L}) \sim {\cal
O}(M^{}_{\rm D} M^{-1}_{\rm R} M^T_{\rm D}) \sim 1~{\rm GeV}$ and
${\cal O}(M^{}_\nu) \sim 0.1~{\rm eV}$ hold and thus the
cancellation can be achieved with the precision of ${\cal
O}(10^{-10})$. It is possible to realize such a fine cancellation in
a more or less natural way by imposing a certain flavor symmetry on
$M^{}_{\rm L}$, $M^{}_{\rm D}$ and $M^{}_{\rm R}$ and introducing
slight perturbations to them \cite{typeII}. In doing so, however,
one has to be very cautious and should take into account the effects
of radiative corrections to two mass terms
\cite{Pilaftsis,Smirnov}.}.
Let us reiterate the key points of our model in the following:
\begin{itemize}
\item We assume that both $M^{}_i$ and $M^{}_\Delta$ are of ${\cal O}(1)$ TeV.
Then the production of $H^{\pm\pm}$ at the LHC is guaranteed, and
their lepton-number-violating signatures will probe the Higgs
triplet sector of the type-II seesaw mechanism. On the other hand,
${\cal O}(M^{}_{\rm D}/M^{}_{\rm R}) \lesssim 10\%$ is possible as a
result of ${\cal O}( M^{}_{\rm R}) \sim 1$ TeV and ${\cal
O}(M^{}_{\rm D}) \lesssim {\cal O}(v)$, such that appreciable
signatures of $N^{}_i$ can be achieved at the LHC.

\item The small mass scale of $M^{}_\nu$ implies that the relation
${\cal O}(M^{}_{\rm L}) \sim {\cal O}(M^{}_{\rm D} M^{-1}_{\rm R}
M^T_{\rm D})$ must hold. In other words, it is the significant but
incomplete cancellation between $M^{}_{\rm L}$ and $M^{}_{\rm D}
M^{-1}_{\rm R} M^T_{\rm D}$ terms that results in the non-vanishing
but tiny masses for three light neutrinos
\footnote{Note that radiative corrections to $M^{}_\nu$ can be
expressed as $\delta M^{}_\nu \approx - (M^{}_{\rm L} \delta^{}_{\rm
L} - M^{}_{\rm D} M^{-1}_{\rm R} M^T_{\rm D} \delta^{}_{\rm R})$,
where $|\delta^{}_{\rm L}| \ll |\delta^{}_{\rm R}| \lesssim 10^{-3}$
is numerically expected in our model. Hence Eq. (6) is a good
approximation irrelevant to small radiative corrections. Although
the magnitude of $\delta M^{}_\nu$ is likely to be much larger than
that of $m^{}_i$ (e.g., of ${\cal O}(1) ~{\rm MeV}$), it can in
principle be suppressed via a kind of more subtle cancellation which
may be accomplished by a slight modification of the relevant Yukawa
couplings. A detailed analysis of radiative corrections to
$M^{}_\nu$ in the TeV-scale type-II seesaw model will be presented
elsewhere.}.
\end{itemize}
In this spirit, $M^{}_{\rm L}$ can be reconstructed via the
excellent approximation $M^{}_{\rm L} = V \widehat{m} V^T + R
\widehat{M} R^T \approx R \widehat{M} R^T$. The elements of the
Yukawa coupling matrix $Y^{}_\Delta$ are then given by
\begin{eqnarray}
\left(Y^{}_\Delta\right)^{}_{\alpha \beta} = \frac{\left(M^{}_{\rm
L}\right)^{}_{\alpha \beta}}{v^{}_\Delta} \approx \sum^3_{i=1}
\frac{R^{}_{\alpha i} R^{}_{\beta i} M^{}_i}{v^{}_\Delta} \; ,
\end{eqnarray}
where the subscripts $\alpha$ and $\beta$ run over $e$, $\mu$ and
$\tau$. This result implies that the leptonic decays of $H^{\pm
\pm}$ depend on both $R$ and $M^{}_i$, which actually determine the
production and decays of $N^{}_i$. Thus we have established an
interesting correlation between the doubly-charged Higgs bosons and
the heavy Majorana neutrinos. To observe the correlative signatures
of $H^{\pm\pm}$ and $N^{}_i$ at the LHC will serve for a direct test
of our type-II seesaw model.

We shall subsequently consider the minimal version of the type-II
seesaw models \cite{minimal}, in which there is only one heavy
Majorana neutrino $N^{}_1$ together with the Higgs triplet $\Delta$,
to illustrate how the collider signatures of $N^{}_1$ and
$H^{\pm\pm}$ are correlated with each other. In this simple but
instructive case, the decay rates of $H^{\pm\pm}$ are given by
\begin{eqnarray}
\Gamma(H^{\pm \pm} \to l^\pm_\alpha l^\pm_\beta) &=& \frac{M^2_1
M^{}_\Delta}{8\pi(1+\delta^{}_{\alpha \beta})v^2_\Delta}
|R^{}_{\alpha 1}|^2 |R^{}_{\beta 1}|^2 \; , ~
\end{eqnarray}
which depend on $M^{}_1$ and $R$ in a very obvious way. Since $R$
determines both the strength of non-unitarity of $V$ and that of
the charged-current interactions of $N^{}_1$, it bridges a gap
between neutrino physics and collider physics.

\section{LHC signatures}

We first look at the production of $N^{}_1$ from proton-proton
collisions at the LHC. The dominant channel is 
$pp \to l^{+(-)}_\alpha N^{}_1 X \to l^{+(-)}_\alpha l^{+(-)}_\beta
W^{-(+)} X$, in which $N^{}_1$ is on-shell produced. Given the
charged-current interactions in Eq. (5), the total cross section of
this process reads
\begin{eqnarray}
\sigma(pp \to l^+_\alpha l^+_\beta W^- X) &\approx& \sigma^{}_{N}
\cdot \frac{|R^{}_{\alpha 1}|^2 |R^{}_{\beta 1}|^2}{\displaystyle 4
\sum_\gamma |R^{}_{\gamma 1}|^2} \; ,
\end{eqnarray}
where $\sigma^{}_{N} \equiv \sigma(pp \to l^+_\alpha N^{}_1
X)/|R^{}_{\alpha 1}|^2$ and the narrow-width approximation has been
used. Note that the reduced cross section $\sigma^{}_{N}$ is
independent of any elements of $R$, and thus it can be computed by
taking account of the parton distribution functions and the mass of
$N^{}_1$. If $N^{}_1$ is much heavier than the gauge bosons and the
SM-like Higgs boson, we have ${\rm Br}(N^{}_1 \to l^+ W^-) \simeq
{\rm Br}(N^{}_1 \to \nu Z^0) \simeq {\rm Br} (N^{}_1 \to \nu h^0)
\approx 25\%$ to a very good degree of accuracy \cite{Pilaftsis}.
That is why there appears a factor $1/4$ on the right-hand side of
Eq. (8).

We proceed to consider the production of $H^{\pm\pm}$ and their
same-sign dilepton events at the LHC. In the narrow-width
approximation, three relevant cross sections can be factorized as
\begin{eqnarray}
\sigma(pp\rightarrow l^+_\alpha l^+_\beta H^- X) & = & \sigma^{}_{H}
\cdot {\rm Br} (H^{++} \to l^+_\alpha
l^+_\beta) \; , \nonumber \\
\sigma(pp\rightarrow l^+_\alpha l^+_\beta W^- X) & = & \sigma^{}_{W}
\cdot {\rm Br} (H^{++} \to l^+_\alpha
l^+_\beta) \; , \nonumber \\
\sigma(pp\rightarrow l^+_\alpha l^+_\beta H^{--} X) & = &
\sigma^{}_{\rm pair} \cdot {\rm Br} (H^{+ +} \to l^+_\alpha
l^+_\beta) \; , ~~
\end{eqnarray}
where $\sigma^{}_{H} \equiv \sigma(pp \to H^{++} H^- X)$,
$\sigma^{}_{W} \equiv \sigma(pp \to H^{++} W^- X)$, and
$\sigma^{}_{\rm pair} \equiv \sigma(pp \to H^{++} H^{--}X)$. The
reduced cross sections $\sigma^{}_{H}$, $\sigma^{}_{W}$ and
$\sigma^{}_{\rm pair}$ only depend on $M^{}_\Delta$ and
$v^{}_{\Delta}$, and they can be calculated in a way similar to the
calculation of $\sigma^{}_{N}$. To detect the
lepton-number-violating signals of $H^{\pm\pm}$, we need to take
account of their decay modes. Because of $M^{}_{H^{\pm \pm}} \approx
M^{}_{H^\pm}$, only two decay modes are kinematically open: $H^{\pm
\pm} \to l^\pm_\alpha l^\pm_\beta$ and $H^{\pm \pm} \to W^\pm W^\pm$
\cite{Han2,Chun}. The leptonic channel is expected to be dominant in
our type-II seesaw model, and a detailed analysis of the branching
ratios of $H^{\pm\pm}$ decays can be found in Ref. \cite{CSXZ}.

To specify the correlation between the signatures of $N^{}_1$ and
$H^{\pm \pm}$ at the LHC, let us parametrize the $3\times 1$
complex matrix $R$ in terms of three rotation angles and three
phase angles \cite{Xing}: $R = (\hat{s}^*_{14}, c^{}_{14}
\hat{s}^*_{24}, c^{}_{14} c^{}_{24} \hat{s}^*_{34})^T$, where
$c^{}_{ij} \equiv \cos \theta^{}_{ij}$ and $\hat{s}^{}_{ij} \equiv
e^{i\delta^{}_{ij}} s^{}_{ij}$ with $s^{}_{ij} \equiv \sin
\theta^{}_{ij}$ (for $ij = 14, 24, 34$). A global analysis of
current neutrino oscillation data and precision electroweak data
yields very stringent constraints on the non-unitarity of the
neutrino mixing matrix $V$, which are equivalent to the
constraints on $R$ \cite{Antusch}: $s^{}_{14} s^{}_{24} \leq 7.0
\times 10^{-5}$ in addition to $s^2_{14}, s^2_{24}, s^2_{34} \leq
1.0 \times 10^{-2}$. On the other hand, the experimental upper
bound on the neutrinoless double-beta decay requires $s^2_{14} (1
+ M^2_1/M^2_\Delta )/M^{}_1 < 5\times 10^{-8} ~{\rm GeV^{-1}}$,
where both the contributions of $N^{}_1$ and $H^{\pm\pm}$ have
been taken into account. Combining all these constraints, we may
choose a typical and self-consistent parameter space of three
mixing angles: $s^{}_{14} \approx 0$, $s^{}_{24} \in [0.01, 0.1]$
and $s^{}_{34} \in [0.01, 0.1]$ \cite{CSXZ}. The decay modes
$H^{\pm\pm} \rightarrow e^\pm e^\pm$, $e^\pm \mu^\pm$ and $e^\pm
\tau^\pm$ are therefore forbidden, while
\begin{eqnarray}
{\rm Br}(H^{\pm \pm} \to \mu^\pm \mu^\pm) & \approx &
\frac{s^4_{24}}{\left(s^2_{24} + s^2_{34}\right)^2} \; ,
\nonumber \\
{\rm Br}(H^{\pm \pm} \to \mu^\pm \tau^\pm) & \approx & \frac{2
s^2_{24} s^2_{34}}{\left(s^2_{24} + s^2_{34}\right)^2}\; ,
\end{eqnarray}
and ${\rm Br}(H^{\pm \pm} \to \tau^\pm \tau^\pm) \approx 1 - {\rm
Br}(H^{\pm \pm} \to \mu^\pm \tau^\pm) - {\rm Br}(H^{\pm \pm} \to
\mu^\pm \mu^\pm)$.
\begin{figure}[t]
  \begin{center}
  \vspace{-1.0cm}
  \includegraphics[height=11.0cm,width=0.49\textwidth]{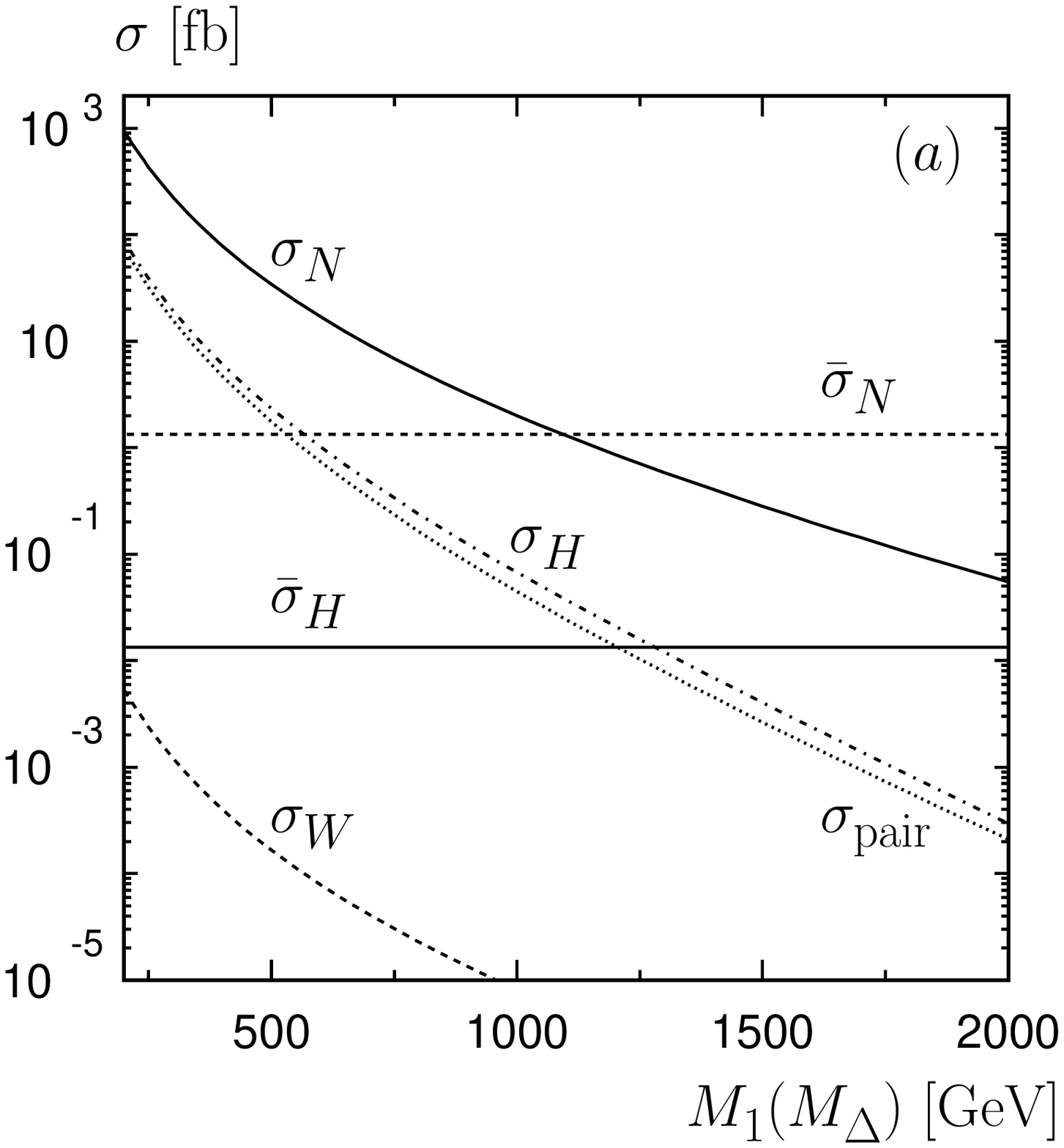}
  \includegraphics[height=11.0cm,width=0.49\textwidth]{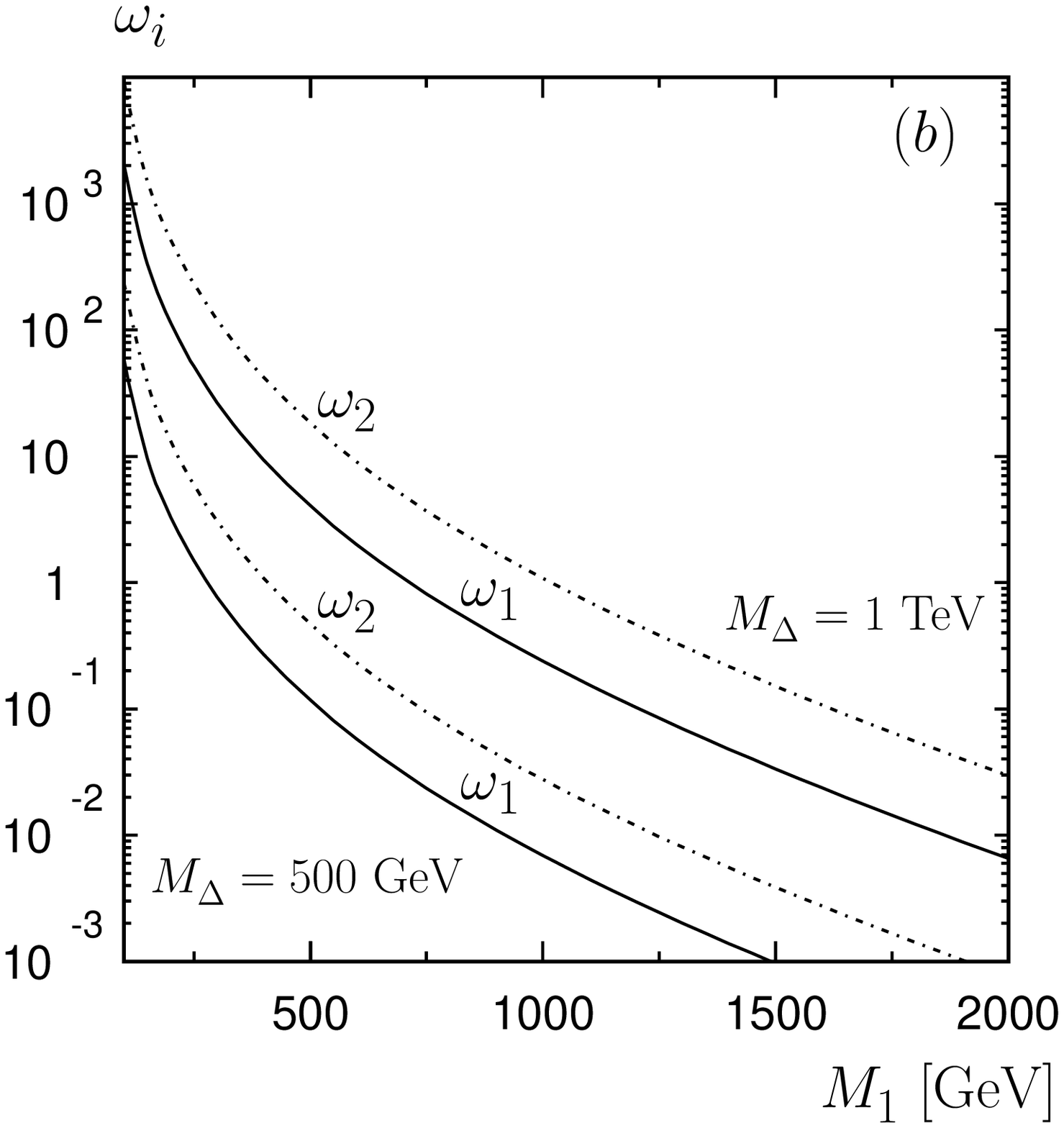}
  \end{center}
  \vspace{-4.5cm}
  \caption{(a) The reduced cross sections $\sigma^{}_{N}$,
$\sigma^{}_{H}$, $\sigma^{}_{W}$ and $\sigma^{}_{\rm pair}$ at the
LHC, where $M^{}_1$ and $M^{}_\Delta$ are the masses of $N^{}_1$ and
$H^{\pm\pm}$, respectively. The horizontal dashed (solid) line
corresponds to the cross section $\bar{\sigma}_{N(H)}$ for one event
induced by heavy Majorana neutrinos (doubly-charged Higgs bosons) at
the LHC with the integrated luminosity of $300~{\rm fb}^{-1}$.
(b) The correlation between the lepton-number-violating signatures
of $N^{}_1$ and $H^{\pm\pm}$ at the LHC, where $s^{}_{14} = 0$,
$s^{}_{24} = s^{}_{34} = 0.1$ and $v^{}_\Delta = 1~{\rm GeV}$ have
typically been input.}
\end{figure}

For each lepton-number-violating process $pp\to l^{\pm}_{\alpha}
l^{\pm}_{\beta} X$ discussed above, its cross section is actually
calculated in the following way:
\begin{eqnarray}
\sigma & = & \sum_{a,b} \int {\rm d}x^{}_1 {\rm d}x^{}_2
F^{}_{a/p}(x^{}_1, Q^2) \cdot F^{}_{b/p}(x^{}_2, Q^2) \cdot
~\hat{\sigma}(ab \to l^\pm_\alpha l^\pm_\beta X) \; ,
\end{eqnarray}
where $F$ denotes the parton distribution function, $x^{}_{1,2}$ is
the energy fraction of the partons, $Q$ is the factorization scale,
and $\hat{\sigma}$ is the partonic cross section. For illustration,
we fix $s^{}_{14} = 0$, $s^{}_{24} = s^{}_{34} = 0.1$, $Q^2 = x^{}_1
x^{}_2 S$ with $\sqrt{S} = 14 ~ {\rm TeV}$, $v^{}_{\Delta} = 1~{\rm
GeV}$ and $ v = 246~{\rm GeV}$ in our subsequent numerical
calculations. We plot $\sigma^{}_{N}$, $\sigma^{}_{H}$,
$\sigma^{}_{W}$ and $\sigma_{\rm pair}$ in FIG. 1 (a) by allowing
$M^{}_1$ and $M^{}_\Delta$ to vary from $200 ~ {\rm GeV}$ to $2 ~
{\rm TeV}$. Note again that these reduced cross sections are
independent of $s^{}_{ij}$. In FIG. 1 (a), we also show the required
cross sections $\bar{\sigma}_{N(H)}$ for one $pp \to \mu^+ \mu^+ X$
event produced at the LHC with the integrated luminosity of
$300~{\rm fb}^{-1}$ in both the case of heavy Majorana neutrino
$N^{}_1$ (dashed line) and that of doubly-charged Higgs boson
$H^{++}$ (solid line). The cross section $\bar{\sigma}^{}_{N(H)}$ is
defined as follows
\begin{equation}
\bar{\sigma}^{}_{N(H)} = \frac{1 ~\rm event}{300~{\rm fb}^{-1}}\;
\frac{1}{B^{}_{N(H)}},
\end{equation}
where
\begin{equation}
B^{}_N =\frac{|R^{}_{\alpha 1}|^2 |R^{}_{\beta 1}|^2}{\displaystyle
4 \sum_\gamma |R^{}_{\gamma 1}|^2}, \; ~~~~~~B^{}_H = {\rm
Br}(H^{++}\to \mu^{+}\mu^{+}).
\end{equation}
 It is obvious that the production rate of $pp \to H^{++} W^-
X$ is remarkably smaller than those of other processes and can be
neglected.
One can find from our results that at the LHC, $H^{\pm \pm}$ (heavy
Majorana neutrinos) should be observable with the integrated
luminosity of $300~{\rm fb}^{-1}$ in the $l^{\pm} l^{\pm}$ channel
up to the mass of $1.2~ {\rm TeV}$ ($1.1 ~{\rm TeV}$). Detailed
analysis is not the aim of this paper and can be found in our
following work. Furthermore, $\sigma^{}_{H}$ is larger than
$\sigma^{}_{\rm pair}$, and the ratio of $\sigma^{}_{H}$ to
$\sigma^{}_{\rm pair}$ lies in the range $1.1 \cdots 1.6$ for
$M^{}_{\Delta} \in [200~{\rm GeV}, 2~{\rm TeV}]$, which is
consistent with the previous results \cite{Han2,Han3}. The
correlation between the LHC signatures of $N^{}_1$ and $H^{\pm\pm}$
becomes more transparent in
\begin{eqnarray}
\omega^{}_1 & \equiv & \frac{\sigma (pp \to \mu^+ \mu^+ W^-
X)|^{}_{N^{}_1}}{\sigma (pp \to \mu^+ \mu^+ H^- X)|^{}_{H^{++}}}
\; , \nonumber \\
\omega^{}_2 & \equiv & \frac{\sigma (pp \to \mu^+ \mu^+ W^-
X)|^{}_{N^{}_1}}{\sigma (pp \to \mu^+ \mu^+ H^{--}
X)|^{}_{H^{++}}} \; ,
\end{eqnarray}
which can approximate to $\omega^{}_1 \approx \sigma^{}_{N}(s^2_{24}
+ s^2_{34})/(4\sigma^{}_{H})$ and $\omega^{}_2 \approx
\sigma^{}_{N}(s^2_{24} + s^2_{34}) /(4\sigma^{}_{\rm pair})$,
respectively. Comparing between Eq. (8) and Eq. (10), we find that
$\omega^{}_{1,2}$ is universal for $\mu \mu$, $\mu \tau$ and $\tau
\tau$ modes. The changes of $\omega^{}_{1}$ and $\omega^{}_2$ with
$M^{}_1$ are illustrated in FIG. 1 (b), where $M^{}_\Delta =
500~{\rm GeV}$ and $1~{\rm TeV}$ have typically been input.

Finally, let us make some brief comments on the type-II seesaw
models with two or three heavy Majorana neutrinos. If the masses of
$N^{}_i$ are nearly degenerate, their production cross sections will
be modified due to either constructive or destructive interference
between different contributions. In particular, the CP-violating
phases of $R$ will enter the expressions of ${\rm Br}(H^{\pm \pm}
\to l^\pm_\alpha l^\pm_\beta)$ and influence the
lepton-number-violating signatures of the doubly-charged Higgs
bosons at the LHC. Because a generic type-II seesaw model contains
many more free parameters even in the case of $M^{}_{\rm L} \approx
R\widehat{M}R^T$, the study of its collider signatures will involve
much more uncertainties. Such an analysis, which is important but
beyond the scope of this paper, will be done elsewhere \cite{CSXZ}.

\section{Summary}

Motivated by the conjecture that new physics at the TeV scale might
not only solve the naturalness problem of electroweak symmetry
breaking but also be responsible for the lepton number (flavor)
violation, we have explored an intriguing possibility to test the
type-II seesaw mechanism of neutrino mass generation at the LHC. Our
key point is that the mass scales of the $SU(2)^{}_{\rm L}$ Higgs
triplet and heavy Majorana neutrinos are both of ${\cal O}(\lesssim
1)$ TeV, and the smallness of $M^{}_\nu \approx M^{}_{\rm L} -
M^{}_{\rm D} M^{-1}_{\rm R} M^T_{\rm D}$ is attributed to a
significant cancellation between its $M^{}_{\rm L}$ and $M^{}_{\rm
D} M^{-1}_{\rm R} M^T_{\rm D}$ terms. This observation allows us to
establish an interesting correlation between the
lepton-number-violating signatures of heavy Majorana neutrinos
$N^{}_i$ and doubly-charged Higgs bosons $H^{\pm \pm}$, but it has
nothing to do with the mass spectrum of three light neutrinos
(either a normal hierarchy or an inverted hierarchy). Taking the
minimal version of the type-II seesaw models for example, we have
calculated the cross sections of $pp \rightarrow l^\pm_\alpha
l^\pm_\beta X$ mediated separately by $N^{}_1$ and $H^{\pm\pm}$, and
illustrated their nontrivial correlation at the LHC. Our results are
quite encouraging, and our analysis can easily be extended to the
more general cases of this class of TeV-scale type-II seesaw models.

We stress that it is extremely important to search for the
correlative collider signatures of $N^{}_i$ and $H^{\pm\pm}$, so
as to convincingly and unambiguously verify the type-II seesaw
mechanism. In contrast, individual signatures of $N^{}_i$ or
$H^{\pm\pm}$ are only possible to demonstrate the type-I or
triplet seesaw mechanism. We also stress that the non-unitarity of
the $3\times 3$ neutrino mixing matrix is intimately correlated
with the LHC signatures of $N^{}_i$. This kind of correlation,
which bridges a gap between neutrino physics at low energies and
collider physics at the TeV scale, deserves a lot of further
investigation.

\section*{Acknowledgment}

One of the authors (Z.S.) is grateful to K. Wang (UW-Madison) for
his help in checking the numerical results. This work was supported
in part by the National Natural Science Foundation of China.

\end{document}